\documentclass[MSNbibl,number,citesort,dvips]{arxstspdf}
\usepackage{flushend}
\usepackage{stfloats}
\volume{28}
\issue{1}
\pubyear{2013}
\firstpage{131}
\lastpage{134}
\doi{10.1214/12-STS389} 


\begin{document}
\begin{frontmatter}

\title{Beno\^{i}t Mandelbrot and Fractional Brownian Motion\thanksref{T1}}
\relateddois{T1}{A French version of this article will appear in
France in the \textit{Gazette des Math\'ematiciens}.}
\runtitle{Beno\^{i}t Mandelbrot and fractional Brownian motion}

\begin{aug}
\author[a]{\fnms{Murad S.} \snm{Taqqu}\corref{}\ead[label=e1]{murad@math.bu.edu}}
\runauthor{M. S. Taqqu}

\affiliation{Boston University}

\address[a]{Murad S. Taqqu is Professor, Department of Mathematics and
Statistics, Boston University, 111 Cummington Street,
Boston, Massachusetts 02215, USA \printead{e1}.}

\end{aug}

%
\begin{abstract}
Although fractional Brownian motion was not invented by Beno\^{i}t
Mandelbrot, it was he who recognized the importance of this random
process and gave it
the name by which it is known today. This is a personal account of the
history behind fractional Brownian motion and some subsequent developments.
\end{abstract}

%
\begin{keyword}
\kwd{Long-range dependence}
\kwd{long memory}
\kwd{self-similarity}
\kwd{Hurst statistic}
\kwd{Beno\^{i}t Mandelbrot}.
\end{keyword}

\end{frontmatter}

\setcounter{footnote}{1}
Since Beno\^{i}t Mandelbrot's passing in October\break 2010, many
well-deserved tributes have been paid to
him.\footnote{There was a special symposium at the {\'E}cole
Polytechnique in Paris in March 2011, one at Yale in April 2011 and a
number of sessions related to Mandelbrot's work took place at the
annual meeting of the American Mathematical Society in Boston in
January 2012.} Beno\^{i}t influenced a great many fields ranging from
the physical sciences to economics, and mathematics was certainly among
them. Beno\^{i}t's great gift was his ability to recognize the hidden
potential in certain mathematical objects.\footnote{Beno\^{i}t
Mandelbrot studied with Paul L{\'e}vy, who is
widely acknowledged for his mastery of the Brownian world.}

I had the good fortune to observe Beno\^{i}t's mathematical analysis in
action, and I would like to tell you about my experience with one of
the objects that Beno\^{i}t worked with, the random process known as a
fractional Brownian motion. Although fractional Brownian motion was
introduced by Kolmogorov, it was Beno\^{i}t Mandelbrot who recognized
the relevance of this random process and, in his seminal paper with Van
Ness \cite{mandelbrotvanness1968}, derived many important
properties. There, he gave this process the name by which it is known
today. See
\cite{mandelbrot2002} for a general review.

Let me recount first how I met Beno\^{i}t. At the beginning of the seventies,
I was a graduate student at Columbia University in the Department of
Mathematical Statistics---a small department but home to prominent
faculty such as Herbert Robbins, David Siegmund and Yuan Shih Chow.
Although I had a fellowship during the academic year, I needed to find
summer work---something I failed to do in my first year. I had sent my
Curriculum Vitae to many companies in New York City, but I did not
receive a single reply.

For my second year, I decided to proceed differently. I asked members
of the Department for contacts. This is how I was put in touch with
Beno\^{i}t Mandelbrot, who was then at IBM Research---an hour's drive
from New York City---but was also nominally an Adjunct Professor in
the Department.
In January of my second year, I called him and inquired about potential
summer jobs. The conversation began in English but quickly turned to
French. I had expected it to last a few minutes, but the conversation
lasted an hour with Beno\^{i}t doing most of the talking (as was often
the case). He ended the conversation, saying that he knew of no jobs.
But a few months later, he called me back. As things developed, it
turned out that he needed a programmer for the summer and asked if I
was interested. I accepted. This is how I became acquainted with his
research at the time, which involved fractional Brownian motion and its
application to hydrology, and how I ended up as Mandelbrot's student.

It started with the so-called ``$R/S$ statistic,'' where $R$ is the
range of partial sums of the data, and $S$ is the sample standard
deviation. It is a statistic that the British hydrologist Harold Edwin
Hurst, in the first half of the twentieth century, had used to study
the yearly variation of the levels of Nile river in Egypt \cite{hurst1951}. The~original work on the subject by Beno\^{i}t Mandelbrot
appeared in 1965
in the Comptes Rendus \cite{mandelbrot1965}.

Under the usual assumptions of finite variance and independent and
identically distributed observations, the $R/S$ statistic should grow
like $n^{1/2}$, where $n$ is the sample size. The Nile data, however,
indicated a growth of $n^H$, where $1/2<H<1$. The growth $n^{1/2}$ is
typically associated with random walk, so $n^H$, with $1/2<H<1$ must
correspond to something else. This is why Mandelbrot suspected that a
process like fractional Brownian motion $B_H(t)$ may perhaps be
relevant in this framework\footnote{Some hydrologists argue instead
that the $n^H$ behavior may be due to nonstationarity. See \cite
{klemes1974} and \cite{montanari2003livre}.} since, while the standard
deviation of Brownian motion at time $t$ is $t^{1/2}$, that of
fractional Brownian motion at time $t$ is $t^H$, where $0<H<1$ \cite
{mandelbrotwallis1968}.
The letter $H$, which refers to the hydrologist Hurst and which was
used by Mandelbrot, has become standard in this context, and it now
labels the fractional Brownian motion.

The term ``fractional Brownian motion'' was coined by Mandelbrot and Van
Ness in the now classical paper \cite{mandelbrotvanness1968}.
Fractional Brownian motion has
a number of nice properties, one of which is ``self-similarity.'' A~process $\{X(t), t\in\mathbb{R}\}$ is self-similar with index $H>0$
if for any $a>0$, the process $\{X(at), t\in\mathbb{R}\}$ has the
same finite-dimensional distributions as $\{a^H X(t),\break t\in\mathbb
{R}\}$. Thus, like a fractal, there is scaling, but it is not the
trajectories of the process that scale, but the probability
distribution, the ``odds.'' This is why this type of scaling is
sometimes called ``statistical self-similarity'' or, more precisely,
``statistical self-affinity.''

The fractional Brownian motion process is then characterized by the
following three properties:
\begin{longlist}
\item[(1)] the process is Gaussian with zero mean;
\item[(2)] it has stationary increments;
\item[(3)] it is self-similar with index $H$, $0<H<1$.
\end{longlist}
Fractional Brownian motion reduces to Brownian motion when $H=1/2$,\vadjust{\goodbreak} but
in contrast to
Brownian motion, it has dependent increments when $H \neq1/2$.
Fractional Brownian motion
was first introduced in 1940 by Andrei Nikolaevich Kolmogorov \cite
{kolmogorov1940}, who was studying spiral curves in Hilbert space. It
was considered by Richard Allen Hunt \cite{hunt1951}
in the context of random Fourier transforms and by Akiva Moiseevich
Yaglom \cite{yaglom1955}, who studied the correlation structure of
processes that have stationary $n$th order increments. However, it is
undoubtedly the seminal paper of Mandelbrot and Van Ness which put the
focus on fractional Brownian motion and gave it its name.
Why the term ``fractional?'' This is because the process can be
represented as an integral with respect to Brownian motion $B(t)$, as follows:
%
\begin{eqnarray}
\label{e:fbm}
\hspace*{20pt}B_H(t) &=& \int_{-\infty}^0 \{(t-s)^{H-1/2} - (-s)^{H-1/2}
\} \,dB(s)\hspace*{-20pt}\nonumber\\[-8pt]\\[-8pt]
&&{}+\int_0^t (t-s)^{H-1/2} \,dB(s)\nonumber\\
&=& \int_{-\infty}^\infty\{{(t-s)}_{+}^{H-1/2} -
{(-s)}_{+}^{H-1/2} \}\, dB(s).
\end{eqnarray}
The integrals are well defined because the integrands are square
integrable with respect to Lebesgue measure. The form of the integrands
is also reminiscent of the one that appears in the $n$-fold iterated
integral formula,
\begin{eqnarray*}
&&\int_0^t dt_{n-1} \int_0^{t_{n-1}} dt_{n-2} \cdots\int_0^{t_2} dt_1
\int_0^{t_1} g(s) \,ds \\
&&\quad=
\frac{1}{(n-1)!} \int_0^t (t-s)^{n-1} g(s) \,ds
\end{eqnarray*}
and therefore (\ref{e:fbm}) can be regarded as involving ``fractional
integrals.'' This, in fact, turns out to be more than a superficial
analogy!

The focus on fractional Brownian motion has proved to be extremely
fruitful because it has allowed all kind of extensions, some of which
were hinted at by Beno\^{i}t Mandelbrot.

For example, the Gaussian noise ``$dB$'' in (\ref{e:fbm}) can be
replaced by an infinite variance L\'evy-stable noise, giving rise to
the linear L\'evy fractional stable motion, which is an infinite
variance self-similar process with stationary, but dependent,
increments \cite{samorodnitskytaqqu1994book}. The kernel can also be
replaced by a random sum of pulses \cite
{cioczekgeorgesmandelbrotsamorodnitskytaqqu1995}. From a different
perspective, the single integral in (\ref{e:fbm})
can be replaced by a multiple integral, so that it becomes an element
of the so-called Wiener chaos
\cite{taqqu1979,peccatitaqqu2011}, of the form
%
\begin{equation}
\label{e:mult}
\int_{\mathbb{R}^k}^\prime g_t (x_1,\ldots,x_k)\, dB(x_1)\cdots\, dB(x_k),
\end{equation}
for a suitable kernel $g_t$ and where prime indicates that one does not
integrate on the diagonals. More specifically, if one chooses
%
\begin{equation}\label{e:hermite-repres-R-2}
g_t (x_1,\ldots,x_k) = \Biggl\{ \int_0^t \prod_{j=1}^k
(s-x_j)_+^{H_0 - {3}/{2}} \,ds \Biggr\},
\end{equation}
where
%
\begin{equation}\label{e:H_0}
H_0 = 1 - \frac{1-H}{k} \in\biggl( 1 -\frac{1}{2k},1 \biggr),
\end{equation}
then the resulting process (\ref{e:mult}) is also self-similar with
index $1/2<H<1$ and has stationary increments. It reduces to fractional
Brownian motion if $k=1$
but is non-Gaussian if $k \geq2$.
The marginal distribution for $k=2$ is studied in \cite{veillettetaqqu2012ros}.

The representation (\ref{e:mult}) with the kernel $g_t$ in (\ref
{e:hermite-repres-R-2})
is called a ``time representation,'' but there are also other
representations, for example, a ``spectral representation'' or a
``finite interval representation'' \cite{pipirastaqqu2010}.

One can also try to define stochastic integrals, where the integrator
is $dB_H$, even though fractional Brownian motion $B_H$ does not have
independent increments. One then needs to define
integrals of the type
%
\begin{equation}
\label{e:int}
\int_{\mathbb{R}} g(x)\, dB_H(x),
\end{equation}
first for nonrandom functions $g$ \cite{pipirastaqqu2002flivre,mishura2008}, and then for random functions $g$ \cite
{biaginihuoksendalhang2008}. One can also consider stochastic
differential equations driven by fractional Brownian motion \cite
{nualart2006}.

In a more applied vein, one can focus on the increments
%
\begin{equation}
\label{e:inc}
X(n) = B_H(n) - B_H(n-1),\quad n \geq1,
\end{equation}
which form a stationary time series with covariance
%
\begin{equation}
r(k) =\mathbb{E} [X(0)X(k)] \sim C k^{2H-2}
\end{equation}
as $k \rightarrow\infty$. The Fourier transform of the covariance
(spectral density),
%
\begin{equation}
f(\lambda) = \sum_{k=-\infty}^\infty r(k) e^{ik\lambda}
\end{equation}
blows up at the origin if $1/2 < H< 1$ since $f(0) = \sum_{k=-\infty
}^\infty r(0) = \infty$.
This type of dependence is called \textit{long memory}, \textit
{long-range dependence} or \textit{strong dependence} \cite{taqqu2003livre}.
Time series with long-memory are important in modeling, particularly in
econometrics. Financial returns, for example, appear uncorrelated, but
their squares often display long-memory \cite{cont2005,andersondaviskreismikosch2009}.

Mandelbrot was very interested in finance and in developing suitable
models for financial returns\vadjust{\goodbreak}
\cite{mandelbrot1997}. Together with his students Adlai Fisher and
Laurent Calvet
\cite{mandelbrotfishercalvet1997},
he introduced a multifractal model of assets returns, $B_H(\theta
(t))$, where $B_H$ is fractional Brownian motion, and $\theta(t)$ is
an independent multifractal process corresponding to ``activity time.''
A~multifractal process has stationary increments and satisfies
%
\begin{equation}
\mathbb{E} |\theta(t)|^q = c(q) t^{\tau(q)+1},\quad t \geq0,
\end{equation}
where $\tau(q)$ is not necessarily a linear function of $q$. Observe
that fractional Brownian motion itself is multifractal, more precisely
monofractal, because by self-similarity, $\mathbb{E} |B_H(t)|^q =
t^{qH}$, corresponding to the linear function $\tau(q) = qH-1$. The
idea of ``subordination,'' replacing ``physical time'' by ``activity
time'' can already be found as early as 1967 in his pioneering paper
with Howard Taylor, with Brownian motion instead of fractional Brownian motion~\cite{mandelbrottaylor1967}.

Statistics also enters into the picture. Given a time series, how does
one check that it displays long-memory? And if it does, how does one
estimate $H$? There is quite a large literature on the subject~\cite
{robinson2003}. Many of the available tests are graphical~\cite
{taqquteverovsky1998on} or asymptotic in nature. The asymptotic ones
are related to central limit theorems but also to so-called noncentral
limit theorems which arise as follows.

Consider a time series $\{ X_n, n \in\mathbb{Z}\} $, and let $s_n^2
= \operatorname{Var} (\sum_{k=1}^n X_k)$. What is the limit of the
normalized sum
%
\begin{equation}
\frac{1}{s_n} \sum_{k=1}^{[nt]} X_k,\quad t \geq0,
\end{equation}
as $n\rightarrow\infty$? It is typically Brownian motion in the case
of weak dependence. But if $\{ X_n, n \in\mathbb{Z}\} $ is
long-range dependent, it could be Brownian motion, fractional Brownian
motion or a non-Gaussian process. The study of limit theorems in this
context is thus very important, and while there is a large literature
about it, there are still many open problems
\cite{taqqu1975,mandelbrot1975L,taqqu1979,dobrushinmajor1979,major1981,embrechtsmaejima2003,nualartpeccati2005}.

This story illustrates one of Beno\^{i}t Mandelbrot's contributions to
mathematics. He tended to focus on a concept or mathematical object
whose importance was not recognized. He studied and developed it---at
times rigorously, at times heuristically---with the frequent
consequence that the object of his attention became the basis of major
subsequent developments. Because he challenged accepted views, many of
his ideas met with initial resistance. Ultimately though, Beno\^{i}t
Mandelbrot's influence on the course of mathematical thinking has been
far-reaching. He will be greatly missed.

\section*{Acknowledgment}
This work was partially supported
by the NSF Grant DMS-10-07616 at Boston University.
Any opinions expressed here are those of the author and do not
necessarily reflect the views of the NSF.

%

\end{document}